\documentclass[11pt]{article}
\usepackage{amssymb}
\usepackage{latexsym}
\usepackage{amsfonts}
\usepackage{graphicx}
\usepackage{amsmath}
\usepackage{amsbsy}
\usepackage{amsthm}
\usepackage{amscd}
\usepackage{dsfont}
\usepackage{makeidx}

\makeindex
\theoremstyle{definition}

\begin{document}

\title{A secret sharing scheme using groups}
\author{Dimitrios Panagopoulos\\
Presefonis 3 N. Iraklio Attikis 14121 Greece\\
\texttt{dpanagop@yahoo.com}}
\date{24 August 2010}
\maketitle

\begin{abstract}
In this paper a secret sharing scheme based on the word problem in
groups is introduced. The security of the scheme and possible
variations are discussed in section 2. The article concludes with
the suggestion of two categories of platform groups for the
implementation of the scheme.
\end{abstract}

\section{Introduction}
The problem of distributing a secret among a group of n persons in
such a way that it can be reconstructed only if at least t of them
combine their shares was solved independently by A. Shamir
\cite{Shamir} and G. Blakley \cite{Blakley} in 1979. During the
recent years several cryptographic methods used group theoretic
machinery (see e.g. \cite{MSU}). In the present article, combining
these two fields, we use group presentations and the word problem in
groups in order to develop a new secret sharing scheme. It's main
advantage to the schemes mentioned before is that it does not
require the secret message to be determined before each individual
person receives his share of the secret.

In the section following the introduction the scheme is introduced.
The article ends with a general discussion about the scheme and some
suggestions concerning the platform groups which could be used for
its implementation.

\section{The scheme}
Suppose that  a binary sequence must be distributed among n persons
in such a way that at least t of them must cooperate in order to
obtain the whole sequence. The secret sharing scheme consists of the
following steps:
\begin{enumerate}
  \item [Step 1] A group G with finite presentation
  $G=<x_1,x_2,\ldots,x_k/\,r_1,\ldots,r_m>$ and soluble word problem is
  chosen. We require that $m=\left(\begin{array}{c}
n\\
t-1
  \end{array}\right)$.
  \item [Step 2] Let $A_1,\ldots,A_m$ be an enumeration of the subsets of
  $\{1,\ldots,n\}$ with t-1 elements. Define n subsets of
  $\{r_1,\ldots,r_n\}$, $R_1,\ldots,R_n$ with $r_j\in R_i$ if and
  only if $i\notin A_j$, $j=1,\ldots,m,\, i=1,\ldots,n$.

  Then for every $j=1,\ldots,m$, $r_j$ is not contained in exactly
  t-1 of the subsets $R_1,\ldots,R_n$. It follows that $r_j$ is
  contained in any union of t of them whereas if we take any t-1 of
  the $R_1,\ldots,R_n$ there exists a j such that $r_j$ is not
  contained in their union.
  \item [Step 3] Distribute to each of the n persons one of the sets
  $R_1,\ldots,R_n$. The set $\{x_1,\ldots,x_k\}$ is known to all of
  them.
  \item [Step 4] If the binary sequence to be distributed is $a_1\cdots a_l$
  construct and distribute a sequence of elements $w_1,\ldots,w_l$ of G such that
  $w_i=_G1$ if and only if $a_i=1$, $i=1,\ldots,l$. The word $w_i$
  must involve most of the relations $r_1,\ldots,r_m$ if $w_i=1$.
  Furthermore, all of the relations must be used at some point in the
  construction of some element.
\end{enumerate}

Any t of the n persons can obtain the sequence $a_1\cdots a_l$ by
taking the union of the subsets of the relations of G that they
possess and thus obtaining the presentation
$G=<x_1,x_2,\ldots,x_k/r_1,r_2,\ldots,r_m>$ and solving the word
problem $w_i=_G1$ in G for $i=1,\ldots,l$.

A coalition of fewer than t persons cannot decode correctly the
message since the union of fewer than t of the sets $R_1,\ldots,R_n$
contains some but not all of the relations $r_1,\ldots,r_m$. Thus
such a coalition could obtain a group presentation
$G'=<x_1,\ldots,x_k/\,r'_1,\ldots,r'_p>$ with $p<m$ and $G\neq G'$,
where $w_i=_G1$ is not equivalent to $w_i=_{G'}1$ in general.

\section{Remarks and implementation}
It should be pointed out that, contrary to other schemes (e.g.
Shamir's, Blakley's scheme), the secret sequence to be shared is not
needed until the final step. It is possible for someone to
distribute the sets $R_1,\ldots,R_n$ and decide at a later time what
the sequence would be. In that way the scheme can also be used so
that t of the n persons can verify the authenticity of the message.
In particular the binary sequence in step 4 could contain a
predetermined subsequence (signature) along with the normal message.
Then t persons may check whether this predetermined sequence is
contained in the encoded message thus validating it. One word of
caution though, such a use might make possible for less than t
persons (or even a third party) to discover all of the relations
$r_1,\ldots,r_m$. This can be made more difficult by not specifying
where exactly the signature should appear.

One method of attack to this system is to search the pool of
possible presentations of groups
$G=<x_1,x_2,\ldots,x_k/\,r_1,\ldots,r_m>$ that are used in the first
step and try to decode the transmitted message $w_1,\ldots w_l$.
This task is easier if the attacker has some information concerning
the encoded message (e.g. the attacker may knows that a certain
block of the message contains a specific binary sequence/singature
as discussed in the previous paragraph). Thus, this pool must
contain a large number of groups. The reader may consult
\cite[6.1.5]{MSU} for further discussion on the efficiency of this
type of attack.

The above line of attack is expedited if the attacker possesses some
of the sets $R_1,\ldots,R_n$ (e.g. he might be one of the n persons
sharing the secret). For this the reason we require in step 4 that a
word $w$ encoding 1 must involve most of the relations. Because if
someone possesses the relations $r'_1,\ldots,r'_p$ and only them are
involved in a word $w=_G1$ then he may decode correctly the word
since $w=_{G'}1$ for the group
$G'=<x_1,\ldots,x_k/\,r'_1,\ldots,r'_p>$.

One way of creating a word representing 1 is by the product
\[\prod_j^l[r'_j,w_j]\]
where $r'_j$ is a relation, $w_j$ a random element, $l$ a (large)
natural number and $[a,b]=aba^{-1}b^{-1}$ is the commutator of a and
b. This kind of encoding might, also, render useless some of the
quotient attacks \cite[6.1.6]{MSU}. A larger set of relations in
step 1 should make these attacks more difficult to use. One the
other hand, the fact that by using only the relationships contained
in $R_j$ for a word $w$ the person with this set can decode
correctly the word, may be used to send messages to a specific
person secretly from the rest of the group.

Finally we propose some categories of group presentations which
could be used in step 1:
\begin{description}
  \item [Polycyclic groups:] polycyclic groups with presentation
  \[<x_1,\ldots,x_k/\,x_j^{a_i}=w_{ij},\,a_j^{a_i^{-1}}=v_{ij},\,a_l^{r_l}=u_l\,\mbox{ for $1\leq i<j\leq k$, $l\in I$}>\]
  where $I\subseteq\{1,\ldots,k\},\,r_l\in\mathds{N}$ for all $l\in
  I$, $w_{ij},v_{ij}, u_j$ are words in $a_{j+1},\ldots,a_k$ and
  $x^y=y^{-1}xy$. The interested reader may consult \cite{EK} for a
  discussion on the use of polycyclic groups.
  \item [Coxeter groups:] Coxeter groups with presentation
  \[<x_1,\ldots,x_k/\,(s_is_j)^{m_{ij}}=1,\,i,j=1,\ldots,k>\]
  where $m_{ij}\in \mathds{N}\cup\{+\infty\}$,
  $m_{ij}\neq0,\,m_{ii}=1$. There exists extensive bibliography on
  Coxeter groups. A place to start is \cite{Cohen}. In there there
  is reference on the word problem in Coxeter groups.
\end{description}

\end{document}